\documentclass[aps,pre,%
preprint,%
amssymb,amsmath,%
nopreprintnumbers,%
showpacs,%
showkeys,%
fleqn,%
eqsecnum,%
byrevtex]{revtex4}
\usepackage{calc}
\usepackage[final]{graphicx}
\begin{document}
\title{The new screening characteristics of strongly non-ideal and dusty
plasmas. Part 1: Single-component systems}
\author{A. A. Mihajlov$^{1}$}
\email{mihajlov@phy.bg.ac.rs}
\author{Y. Vitel$^{2}$}
\email{yv@ccr.jussieu.fr}
\author{Lj. M. Ignjatovi\'c$^{1}$}
\email{ljuba@phy.bg.ac.rs}

\affiliation{$^{1}$Institute of Physics, P.O. Box 57, 11080 Zemun,
Belgrade, Serbia} \affiliation{$^{2}$Laboratoire des Plasmas Denses,
Universite P. et M. Curie, 3 rue Galilee, Paris, 94200 Ivry sur
Seine, France}
\begin{abstract}
In this paper a new model method for describing of the electrostatic
screening in single-component systems which is free of
Debye-H\"{u}ckel's non-physical properties is presented. The method
is appropriate for the determination of screening parameters in the
case of the systems of higher non-ideality degree. The obtained
screening characteristics are presented in a simple analytic form.
The presented results make basic elements of a method for
determination of screening characteristics in dense two-component
plasmas, which are discussed in Part2 and Part 3 of this work.
\end{abstract}
 \maketitle
\section{Introduction}
\label{sec:intro}

It is well known that, beside the existing strict methods, model
methods are also used in all parts of plasma physics and in
connection with most diverse problems. The purpose of every model
method is to illuminate the real meaning of the problem considered
in a physically acceptable, but much simpler way in comparison with
the corresponding strict methods
\cite{suc64,deu77,deu78,gun83,gun84,kra86,mih89,kob95,ram03}.
Probably the best known model methods is Debye-H\"{u}ckel's (DH
method), which was developed for describing electrostatic screening
in electrolytes \cite{deb23}, but now for a long time it has been
used in the plasma physics. Namely, a lot of electrostatic screening
effects in plasmas are very often interpreted just in terms of such
products of DH method as Debye-H\"{u}ckel's screening potential (DH
potential), as well as Debye's radius and screening constant
\cite{spi62,gri64,dra65,ich73,gri74,ebe76,kra86}. Because of that
one could find these products of DH method in every course of plasma
physics.

When talking about DH method it is usual to include in this concept
also the corresponding basic model of electrostatic screening in the
considered system. According to \cite{deb23} that model understands:
\begin{itemize}
\item (a1) the presence of an immobile probe particle, which
represents one kind of charged particles in the real system (plasma,
electrolyte);
\item (a2) treatment of free charged particles which fill the space around
the probe particle as an ideal gas in the state of thermodynamical
equilibrium;
\item (a3) treatment of the average summary electrostatic field as external
with respect to that ideal gas.
\end{itemize}
However, here we will adopt an agreement that in the further
consideration "DH method" means exclusively the way of usage of the
basic model, while this model itself will be considered
independently. Namely, although the basic model was introduced in
\cite{deb23} together with DH method, this beneficent model has its
own significance, since it can be used for development of another
method for describing of electrostatic screening. In \cite{deb23}
the basic model was applied in a showy way which provided for DH
method to possess the following very attractive features:
\begin{description}
  \item (b1) - the procedure used is self-consistent, which means that all
the existing quantities are determined within the procedure itself
and expressed through the basic plasma parameters (electron density
$N_{e}$, temperature $T$, etc.),
  \item (b2) - the final results are presented by simple and compact
analytical expressions.
\end{description}
Certainly, it were these features that maintained the constant
popularity of this method, although strict approaches based on
classical or quantum statistical mechanics also exist
\cite{ich73,ebe76,kra86}.

However, the features (b1) and (b2) prevent from noticing the fact
that DH method has at least two serious non-physical properties,
masked by the applied mathematical procedure, which is discussed
within this work. In this context one should remind to the problem
connected with the screening constant for two-component systems.
Namely, it was noticed that instead of DH constants, referring to
the real two-component systems, the constants referring to the
corresponding one-component systems were often used. Such a
situation was probably first announced in \cite{bow61}, where DH
screening constant of electron gas (on a positive charged
background) was used instead of the one referring to the considered
gaseous plasma. Later, gaseous plasma was treated in a similar way
in several cases \cite{dra65,vit90,vit01}, as well as in the cases
of very dense non-ideal plasmas
\cite{ada80,dju91,ada94,mih01,ada04}. This reflects the fact that
one of mentioned non-physical properties manifests in every
multi-component system independently of its non-ideality degree. It
means that {\it in the case of multi-component systems DH method in
principle gives wrong results, even if the non-ideality degree is
very small}. Consequently, it was natural to expect that a new model
method, free of all non-physical properties, but keeping positive
features (b1) and (b2), would be more appropriate for plasmas of
higher non-ideality degree, including dusty plasmas too. This was
one of the main stimuli for this research whose purpose to develop a
new model method which would be more adequate for description of
inner-plasma electrostatic screening.

An other stimulus was the fact that in the laboratory practice for a
long time there has been {\it a need for finding new screening
characteristics of dense non-ideal plasma} ($N_{e} = 10^{18} \div
10^{22}cm^{-3}$, $T = 1.5\cdot 10^{4} \div 3\cdot 10^{4}K$) {\it
which would be different from Debye's radius and screening
constant.} It is illustrated by many papers
\cite{kak73,gun83,gun84,mih89,vit90,shv98,vit01}, where such new
screening characteristics, introduced semi-empirically, were
discussed. Besides, in some other papers such new characteristics
were also used, although without any special discussion
\cite{ada80,dju91,ada94,mih01,ada04}. Because of that, a need was
evident for a new model method which should be the generator of some
screening characteristics appropriate for interpretation of
experimental data.

Already at the beginning of this research it was found that it is
not possible to develop the searched model method by means of any
corrections to DH one, and a completely new approach has to be
found. However, we started from the analysis of the same basic
model, keeping in mind that its exceptional properties (a1)-(a3)
allow including the inner-plasma electrostatic screening into
consideration in the simplest way, and expecting that the model
could offer some new possibilities. During this research it was
found that {\it DH method does not exhaust all the possibilities of
the basic model}, and that {\it this model allows construction of a
new method which posses positive characteristics (b1) and (b2), but
which is free of all the negative properties of DH one}. The main
aim of this work is just the presentation of this method, as well as
several new screening characteristics obtained by its means.

Due to the huge amount of problems which have to be studied in the
general case of multi-component systems, only two simplest cases,
i.e. the cases of single- and two-component systems, were considered
in this work. The whole material has been arranged in three papers.
The new model methods for single- and two-component systems are
developed in Part 1 and Part 2, respectively, and the obtained new
screening characteristic lengths and non-ideality parameters are
presented in Part 3. The comparison of the results obtained by the
developed method with the existing experimental data is also
performed in Part 3.

As the main objects of our researches we treated such two-component
systems as fully ionized non-ideal gaseous plasmas (before of all
hydrogen and helium ones), dusty plasmas, some electrolytes etc.
However, in Part 1 we start just from single-component systems (an
electron gas on the positive background etc.), since in that case
the basic model itself is much simpler than in two-component case.
Because of that, the procedure of elimination of one of the
mentioned non-physical properties, which appears in all considered
systems, can be developed in the simplest way. Consequently, in Part
2 it will be possible to concentrate whole attention to the
elimination of other non-physical property of DH method. Also, we
kept in mind the fact that a single-component system may always be
treated as an approximation of a multi-component system. Therefore,
the parameters of the corresponding single-component system could
serve as an estimation in advance of the parameters of the observed
multi-component system.

The material presented in Part 1 is distributed in the next three
Sections, as well as in the three Appendixes. Section \ref{sec:ass}
contains: description of the screening model, a critical analysis of
DH method in the case of single-component system and stating the
tasks precisely. In Section \ref{sec:mdm} the method developed for
the single-component system is presented. Finally, Section
\ref{sec:rd} contains results and discussion.

\section{Theory assumptions}
\label{sec:ass}

\subsection{Screening model}
\label{sec:commass}

A stationary homogeneous single-component system $S_{in}$ is taken here as
the initial model of some real physical objects. We will assume that:
$S_{in}$ is constituted by a gas of charged particles of only
one kind and a non-structured charged background; the gas there is
in an equilibrium state with temperature $T$ and mean local particle
density $N$; the particles are point objects with charge $Ze$, where
$Z = \pm 1, \pm 2$, etc., and $e$ is the modulus of the electron charge;
the parameters $Z$, $N$ and the background charge density $\rho_{b}$
satisfy the local quasi-neutrality condition
\begin{equation}
\label{eq1}  \rho_{b} + Ze \cdot N = 0.
\end{equation}
Also, in the case of the electron gas it is assumed that the values
of $N$ and $T$ allow its non-relativistic treatment.

Electrostatic screening of a charged particle in the system $S_{in}$
is modeled with the help of the corresponding accessory system
$S_{a}$ which, accordingly to the basic model properties (a1) and
(a2), contains: a probe particle with the charge $Ze$, fixed in the
origin of chosen reference frame (the point $O$),
the charged background identical to that in $S_{in}$, and the gas of the free charged
particles in the state of thermodynamical equilibrium with the same $Z$ and $T$
as in $S_{in}$.

The gas of free particles in the system $S_{a}$ will be
characterized by: the mean local particle density $n(r)=n(r;Z,T)$,
the mean total charged density
\begin{equation}
\label{eq2}  \rho (r) = \rho_{b} + Ze \cdot n(r),
\end{equation}
and the mean electrostatic potential $\Phi (r)$, where $r=|\vec{r}|$
and $\vec{r}$ is the radius-vector of the observation point. It is
presumed that $n(r)$ and $\rho(r)$ satisfy the asymptotic boundary
condition
\begin{equation}
\label{eq3}  \lim\limits_{r \to \infty } n(r) = N
\end{equation}
and the condition of neutrality of the system $S_{a}$ as a whole
\begin{equation}
\label{eq4}  Ze + \int\limits_0^\infty {\rho (r)} \cdot 4\pi r^2dr =
0.
\end{equation}
Then, we will take into account that $\Phi(r)$ and $\rho(r)$ are
connected by Poisson's equation
\begin{equation}
\label{eq5}  \nabla^{2} \Phi = - 4\pi [Ze \cdot \delta(\vec{r}) +
\rho (r)],
\end{equation}
where $\delta(\vec{r})$ is the three-dimensional $\delta$-function
\cite{ich73}. It is presumed the satisfying of the boundary
conditions
\begin{equation}
\label{eq6}  \lim\limits_{r \to \infty } \Phi (r) = 0,
\end{equation}
\begin{equation}
\label{eq7}  \vert \varphi \vert < \infty , \qquad \varphi \equiv
\mathop {\lim }\limits_{r \to 0}[\Phi (r) - Ze/r].
\end{equation}
Since $\varphi$ is the mean electrostatic potential in the point
$O$, the quantity
\begin{equation}
\label{eq8}  U = Ze \cdot \varphi
\end{equation}
is the mean potential energy $U$ of the probe particle. In an usual
way we will treat $U$ as an approximation of the mean potential
energy of a free charged particle in the system $S_{in}$.

In accordance with the basic model properties (a2) and (a3), as well
as the boundary conditions (\ref{eq3}) and (\ref{eq6}), the
condition of the keeping of thermodynamical equilibrium in the
system $S_{a}$ can be presented in the form
\begin{equation}
\label{eqmu}  \mu \left({n(r),T} \right)+Ze\Phi (r)=\mu \left( {N,T}
\right),
\end{equation}
where $\mu (n,T)$ has a sense of the chemical potential of the ideal
gas in the state of thermodynamical equilibrium with the particle
density $n$ and temperature $T$. Within this model just this
equation is another one which, together with Eq.~(\ref{eq5}),
provides the determination of charge density $\rho(r)$ and the
potential $\Phi(r)$. Let us emphasize that Eq.~(\ref{eqmu}) is
applicable not only to the classical systems, but to the degenerated
systems too (see \cite{ebe76,kit77}). However, one should keep in
mind the difference between Eqs.~(\ref{eq5}) and (\ref{eqmu}).
Namely, while Eq.~(\ref{eq5}) is applicable in the whole space,
Eq.~(\ref{eqmu}) is valid only in the area where usage of chemical
potential $\mu (n(r),T)$ has a physical meaning. In the considered
case, when the probe particle and free particles have the same
charge, Eq.~(\ref{eqmu}) is principally valid only in the region
\begin{equation}
\label{eqrrs} r \gtrsim r_{s}, \qquad r_s \equiv \left(\frac{3}{4\pi N}\right)^{1/3},
\end{equation}
where $r_{s}$ is Wigner-Seitz's radius for the system $S_{in}$.
Below $r_{s}$ will be treated as the probe particle self-sphere.

In further considerations we will use the fact that equation
(\ref{eqmu}) can be presented in the linearized form
\begin{equation}
\label{eq9}  \displaystyle{n(r) - N = \frac{Ze}{\partial \mu
/\partial N} \cdot \Phi (r)}, \quad \partial \mu/\partial N \equiv
\left[ \frac{\partial \mu (n,T)}{\partial n} \right]_{n = N}
\end{equation}
in the region of $r$, where
\begin{equation}
\label{eq12}  \frac{ \left| {n(r) - N} \right|}{N} \ll 1.
\end{equation}
Accordingly to the boundary condition Eq.~(\ref{eq3}) the part of
space where the condition (\ref{eq12}) is satisfied always exists.

\subsection{The critical analysis of DH method}
\label{sec:kritD}

\begin{figure}[htbp]
\centerline{\includegraphics[width=\columnwidth,
height=0.75\columnwidth]{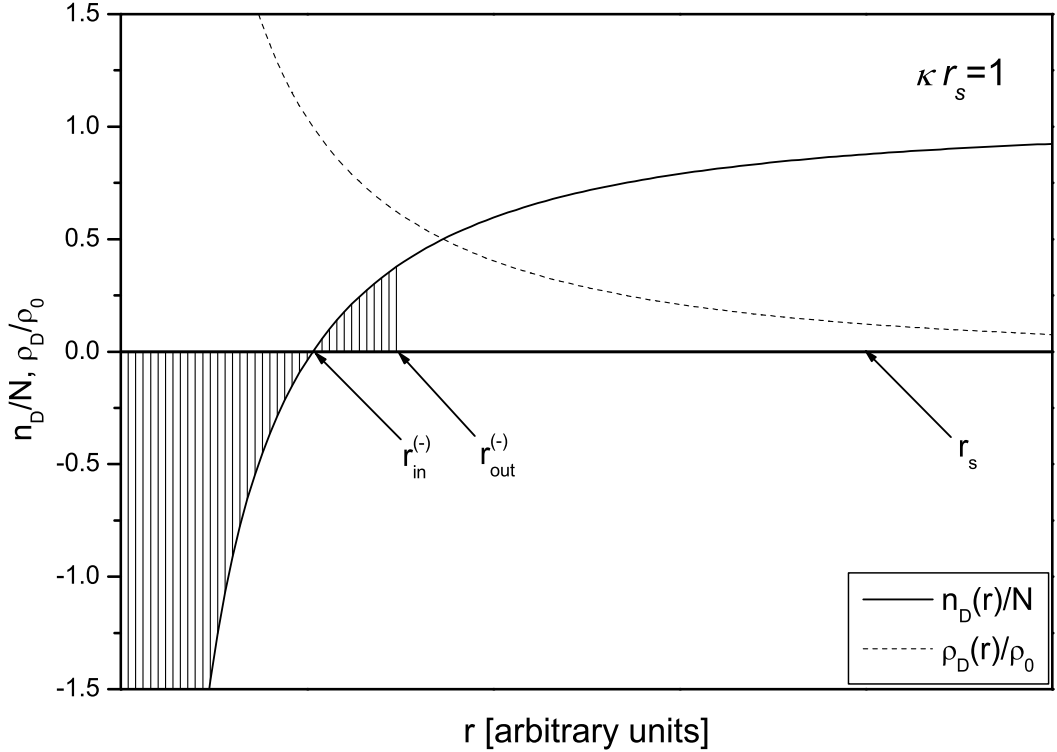}} \caption{The reduced particle
density $n_D (r) / N$ and the reduced charge density $\rho _D (r) /
\rho_{b} $ in the case $\kappa r_s = 1$, where $\kappa $ is Debye's
screening constant given by (\ref{eq18}).} \label{fig:nrhoD}
\end{figure}

The procedure of obtaining of DH solutions $\Phi _D (r)$, $\rho _D
(r)$ and $n_D (r)$ for the electrostatic potential and the charge
and particle densities in the single-component case, as well as
their properties, are described in Appendix \ref{sec:appDebye}. The
behavior of the reduced particle and charge densities, $n_D (r) / N$
and $\rho _D (r) / \rho_{b} $ for one typical case is illustrated by
Fig.~\ref{fig:nrhoD}. This figure demonstrate apparent disadvantages
of DH method: the negativity of the solution $n_D (r)$ in the region
$r < r_{in}^{( - )}$ with the singularity in the point $O$; the
existence of an additional non-physical region $r_{in}^{( - )} <
r_{out}^{( - )} < r_{s}$, whose the sole role within DH method is to
compensate the influence of the region $r < r_{in}^{(-)}$. The
distances $r_{in}^{( - )}$ and $r_{out}^{( - )}$ are the roots of
the equations (\ref{eqnula1}) and (\ref{eqnula2}), which always
satisfy non-equalities $0< r_{in}^{( - )} < r_{out}^{( - )} <
r_{s}$. Apart of that, in Appendix \ref{sec:appDebye} the attention
is driven to the fact that for strongly non-ideal systems the direct
manifestations of the non-physically behavior of $n_D (r)$ appear
also in the region $r > r_{s}$.

One can see that all non-physical properties of DH solutions are
caused by the used procedure itself, which requires obtaining of the
potential $\Phi_{D}(r)$ as first. In this procedure $\Phi_{D}(r)$ is
the solution of Helmholtz's equation (\ref{eqHelm}) determined in
the whole space, under  the conditions (\ref{eq6}) and (\ref{eq7}),
since in the considered case the natural boundary conditions exits
only $r = \infty$ and $r = 0$. In such a way it is completely
neglected the fact that Eq.~(\ref{eqHelm}) was obtained by means the
equation (\ref{eq9}) which is not applicable inside the probe
particle self-sphere ($0 < r < r_{s}$).

\subsection{What one should do in order to build a physically correct
method?} \label{sec:nonD} In accordance with above mentioned our
main task will be the finding (within the basic model) of procedure
which would provide satisfaction of the non-negativity condition
\begin{equation}
\label{eq10}  n(r) \ge 0, \qquad 0 < r < \infty,
\end{equation}
as well as the application of the equation (\ref{eq9}) only under
the condition (\ref{eq12}). Also, we will keep in mind the fact that
the basic equation (\ref{eqmu}) is correct only in the region
(\ref{eqrrs}), wherefrom it follows that {\it in the region} $r <
r_{s}$ {\it the searched procedure mast not base on the any equation
obtained from (\ref{eqmu}), including here the equation
(\ref{eq9})}. It is clear that, contrary to DH procedure, the first
aim of the searched procedure has to be determination of the free
particle density $n(r)$.

\section{The method presented}
\label{sec:mdm}

\subsection{The solution $n(r)$}

{\bf The region of large $r$: alternative procedure.} In accordance
with above request we will use Poisson's equation (\ref{eq5}) in
order to obtain the potential $\Phi (r)$ for given charge density
$\rho(r)$. For this purpose we will use the expression (\ref{eqB3}).
In the form (\ref{eqB3}) the potential $\Phi (r)$ will be taken in
the equation (\ref{eq9}). After the multiplication by $Ze$, it
transforms in the integral equation of Volterra's type, namely
\begin{equation}
\label{eq17}  \rho (r) = \kappa ^2\int\limits_r^\infty {\rho
(r')\left( {\frac{1}{r} - \frac{1}{r'}} \right)} r'^2dr',
\end{equation}
where the screening constant $\kappa$ is given by the expression:
\begin{equation}
\label{eq18}  \kappa \equiv \frac{1}{r_\kappa } = \left[ \frac{4\pi
(Ze)^2}{\partial \mu/\partial N} \right]^{1 \over 2},
\end{equation}
and $r_{\kappa}$ is the one of the characteristic length which
appear in considered model. Let us draw attention that $\kappa =
\kappa _D $ in the classical case ($\partial \mu / \partial N = kT /
N)$, and $\kappa = \kappa _{T - F} $ in the case of ultra
degenerated electron gas $(T = 0,\,\,\,\partial \mu /
\partial N = 2\varepsilon _F / 3N)$, where
\begin{equation}
\label{eq19}  \kappa _D \equiv \frac{1}{r_D } = \left[ {\frac{4\pi
(Ze)^2}{kT}N} \right]^{1 \over 2}, \mbox{} \kappa _{T - F} \equiv
\frac{1}{r_{T - F} } = \left( {\frac{6\pi }{\varepsilon _F }N}
\right)^{1 \over 2},
\end{equation}
and $r_D $ and $r_{T - F} $ are known Debye's and
Thomas-Fermi radii, and $\varepsilon _F $ is the corresponding Fermi
energy (see \cite{ebe76,kit77}). The principal significance of usage
of just described procedure is caused by the fact that for any $r$
in (\ref{eq17}) only the region $r < r' < \infty$ appears. Because
of that (\ref{eq17}) can be treated within the region of $r$ where
the condition (\ref{eq12}) is satisfied.

The solution of the equation (\ref{eq17}) will be found in the form:
$\rho (r) = S(r)/r$. After that (\ref{eq17}) gets the form
\begin{equation}
\label{eq20}  S(r) = \kappa ^2\int\limits_r^\infty {S(r')(r' -
r)dr'}.
\end{equation}
In order to determine $S(r)$ we will apply the operator $d^2/dr^2$
to both the left and the right side of (\ref{eq20}) and obtain the
equation
\begin{equation}
\label{eq21}  \frac{d^2S(r)}{dr^2} = \kappa ^2S(r).
\end{equation}
The general solution of this equation is $S(r) = A\exp ( - \kappa r)
+ B\exp (\kappa r)$. Since (\ref{eq20}) is satisfied only in the
case $B = 0$, we have it that: $\rho (r) = A \cdot r^{-1}\exp ( -
\kappa r)$. On the base of this and (\ref{eq2}) we obtain, taking $A
= - Ze \cdot a$, the relation
\begin{equation}\label{eq22a}
 n(r) = N - a  \cdot \frac {\exp ( - \kappa r)}{r},
\end{equation}
where is taken that $a > 0$. One can see that the alternative
procedure, which is used here, provides analyzing of solution $n(r)$
in the region where it satisfies the condition (\ref{eq12}). The way
of determination of the coefficient $a$ in Eq.~(\ref{eq22a}) and
obtaining $n(r)$ in the whole region $0 < r < \infty$ is described
in details in Appendix \ref{sec:appsolutionn}.

\begin{figure}[htbp]
\centerline{\includegraphics[width=\columnwidth,
height=0.75\columnwidth]{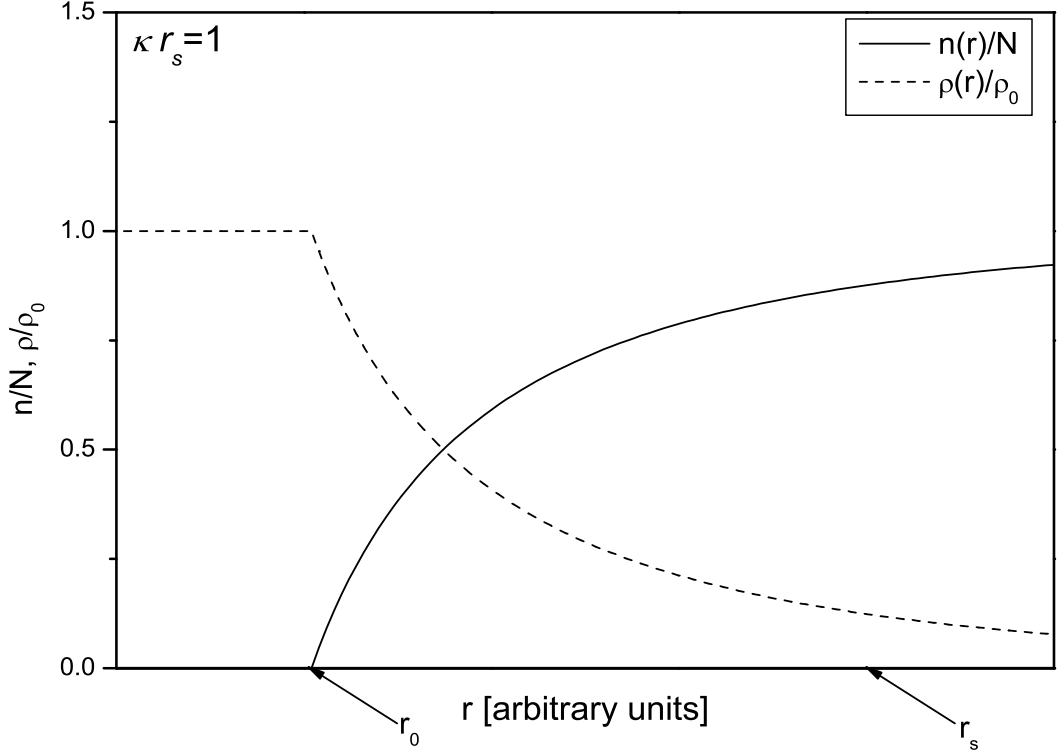}} \caption{The reduced particle
density $n(r) / N$ and reduced charge density $\rho (r) / \rho_{b} $
in the case $\kappa r_s = 1$.} \label{fig:nrho}
\end{figure}

{\bf The complete expression.} From (\ref{eq22a}), (\ref{eq22c}) and
(\ref{eq22d}) it follows that the complete solution $n(r)$ is given
by expression
\begin{equation}
\label{eq23}  n(r) = \left\{ {{\begin{array}{*{20}c} \displaystyle{
{N - N\cdot r_0 \cdot \exp (\kappa r_0 ) \cdot \frac{\exp ( - \kappa
r)}{r},}}
\hfill & {r > r_0 ,} \hfill \\
 {0,} \hfill & {0 \le r \le r_0 ,} \hfill \\
\end{array} }} \right.
\end{equation}
where the parameter $r_{0}$ can be determined from the condition
(\ref{eq4}) taken in the form (\ref{eq4a}). In order to determine
the radius $r_0 $ we will take into account that Eq.~(\ref{eq4}),
after dividing with $Ze$, gives the equation
\begin{equation}\label{eq4a}
 \int\limits_0^\infty {\left[ {N - n(r)} \right] \cdot 4\pi r^2}
dr = 1,
\end{equation}
which is especially discussed in Part 3. In the case when $n(r)$ is
given by Eq.~(\ref{eq23}), from (\ref{eq4a}) it follows the equation
\begin{equation}\label{4b}
 \left( {1 + \kappa r_0 } \right)^3 = 1 + (\kappa r_s )^3,
\end{equation}
whose solution can be presented in two equivalent forms, namely
\begin{equation}
\label{eq26}  r_0 = r_s \cdot \gamma _{s}(x) , \qquad r_0 = r_\kappa
\gamma _{\kappa }(x),
\end{equation}
\begin{equation}
\label{eq25}  \gamma _{s}(x) = [(1 + x^3)^{\textstyle{1 \over 3}} -
1]/x, \qquad \gamma _{\kappa }(x) = (1 + x^3)^{1 \over 3} - 1,
\end{equation}
where the parameter $x$ is defined by relations
\begin{equation}
\label{eq27}  x = \kappa r_s = r_s/r_{\kappa }.
\end{equation}
From these expressions it follows that
\begin{equation}
\label{eq27a}  0 < r_{0} < r_{s}
\end{equation}
in the whole region $0 < x < \infty$, and that
\begin{equation}
\label{eq27b}  \lim_{x\to 0} r_{0}=0, \qquad \lim_{x\to \infty} r_{0}=r_{s}.
\end{equation}
Consequently, wigner-Seitz's radius $r_{s}$ represents the upper
boundary for the radius $r_{0}$. The physical meaning of the
coefficients $\gamma _{s }$ and $\gamma _{\kappa }$ is discussed in
Part 3. The behavior of the solutions $n(r)$ and $\rho (r)$ is
illustrated by Fig.~\ref{fig:nrho} for $\kappa r_s = 1$.

\subsection{The solutions $\rho (r)$, $\Phi (r)$ and \\ the potential energy
$U$}
\label{sec:tt}

The expression for the solution $\rho(r)$ is obtained by means of
(\ref{eq1}), (\ref{eq2}) and (\ref{eq23}), and it is presented here
in the form
\begin{equation}
\label{eq24}  \rho (r) = \left\{ {{\begin{array}{*{20}c}
\displaystyle{ {-Ze N\cdot r_0 \cdot \exp (\kappa r_0 ) \cdot
\frac{\exp ( - \kappa r)}{r},}} \hfill &
{r > r_0 ,} \hfill \\
 {-Ze N,} \hfill & {0 \le r \le r_0 ,} \hfill \\
\end{array} }} \right.
\end{equation}
where $r_{0}$ is given by Eqs. (\ref{eq26}) and (\ref{eq27}). The
corresponding expression for $\Phi(r)$, obtains by means of
(\ref{eqB3}) and (\ref{eqB4}) in the form
\begin{equation}
\label{eq23App}  \Phi(r) = \frac{Ze}{r} \cdot \left\{
{{\begin{array}{*{20}c} \displaystyle{ {\exp ( - \kappa r) \cdot
\chi (x),}}
\hfill & {r > r_0 ,} \hfill \\
\displaystyle{1+\frac{\varphi r}{Ze} +
\frac{1}{2}\left(\frac{r}{r_{s}}
\right)^{3},} \hfill & {r \le r_0 ,} \hfill \\
\end{array} }} \right.
\end{equation}
where the factor $\chi (x)$ is given by expression
\begin{equation}
\label{eq29}  \chi (x) = 3\frac{\left( {1 + x^3} \right)^{1 \over 3}
- 1}{x^3} \cdot \exp \left[ {\left( {1 + x^3} \right)^{1 \over 3} -
1} \right].
\end{equation}
The behavior of $\chi (x)$ is illustrated by Fig.~\ref{fig:chi} and
discussed in Appendix \ref{sec:appsolutionn}. The point $x =
7^{\frac{1}{3}}$ is shown in this figure since in this point
$r_{0}=r_{\kappa}$.
\begin{figure}[htbp]
\centerline{\includegraphics[width=\columnwidth,
height=0.75\columnwidth]{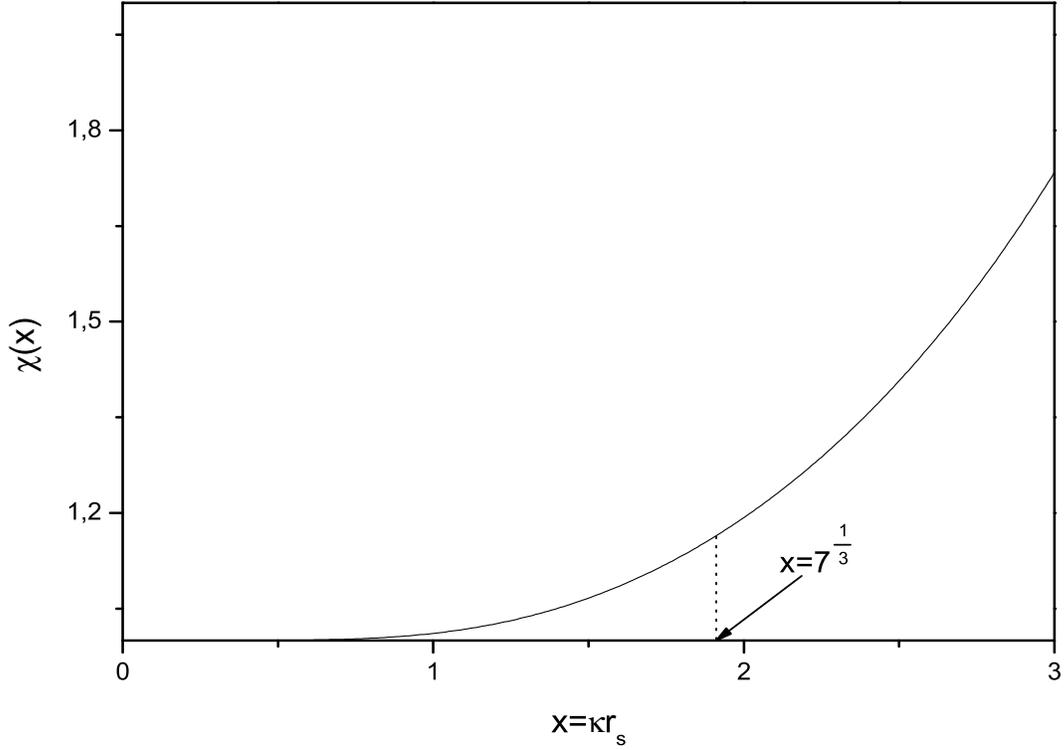}} \caption{The behavior of the
function $\chi (x)$, defined by (\ref{eq29}). In the point $x =
7^{\frac{1}{3}}$ the equality $r_{0}=r_{\kappa}$ is valid.}
\label{fig:chi}
\end{figure}

\begin{figure}[htbp]
\centerline{\includegraphics[width=\columnwidth,height=0.75\columnwidth]
{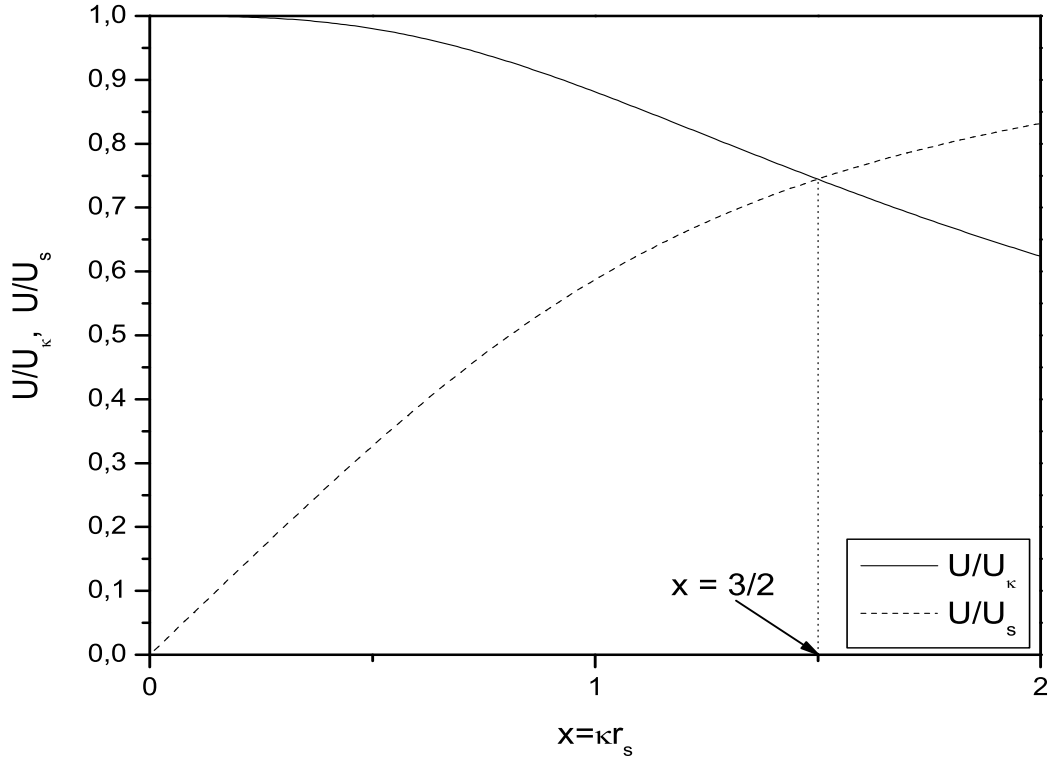}} \caption{The behavior of the ratios $U/U_{\kappa}$
and $U/U_{s}$ as functions of $x$. The point $x = 3/2$ represent an
arbitrary boundary between the regions of weak and strong
non-ideality.} \label{fig:rkrpot}
\end{figure}

Accordingly to Eq.~(\ref{eq8}) determination of the potential energy $U$
of the probe particle requires knowledge of the potential $\varphi$ defined by
Eq.~(\ref{eq7}). By means of Eqs.~(\ref{eq24})-(\ref{eq27}) and Eq.~(\ref{eqB2})
one obtains that
\begin{equation}\label{eqB2App}
 \varphi = -Ze \cdot \frac{3r_{0}}{2r_{s}^{3}}
\left(r_{0}+\frac{2}{\kappa} \right).
\end{equation}
From here and Eqs.~(\ref{eq25}) it follows that  the potential energy $U$
can be presented in two equivalent forms
\begin{equation}
\label{eq30a}  U =  U_{\kappa} \cdot {\frac{3}{2}\frac{\left( {1 +
x^3} \right)^{2 \over 3} - 1}{x^3}} , \qquad U_{\kappa} \equiv
-\frac{(Ze)^{2}}{r_{\kappa}},
\end{equation}
\begin{equation}
\label{eqUs}  U = U_{s} \cdot {\frac{\left( {1 + x^3} \right)^{2
\over 3} - 1}{x^2}}, \qquad U_{s}\equiv -\frac{3}{2} \cdot
\frac{Ze}{r_{s}},
\end{equation}
where $U_{\kappa}$ is DH potential energy of the probe particle (see
Appendix \ref{sec:appDebye}), and $U_{s}$ is the potential energy of
this probe particle only in the field of the charged background
which fills its self-sphere ($0 < r \le r_{s}$). The parameters
$U_{\kappa}$ and $U_{s}$ represent the corresponding boundary
potential energies, since from (\ref{eq30a}) and (\ref{eqUs}) it
follows that
\begin{equation}
\label{eqrpot2}   \lim\limits_{x \to 0}U/U_{\kappa} = 1, \qquad
\lim\limits_{x\to \infty}U/U_{s} = 1.
\end{equation}
The behavior of the the ratios $U/U_{\kappa}$ and $U/U_{s}$ is
illustrated by Fig.~\ref{fig:rkrpot}.

\section{Results and discussion}
\label{sec:rd}

The expressions (\ref{eq23})-(\ref{eq27}) and
(\ref{eq23App})-(\ref{eqUs}) show that the obtained solutions
$n(r)$, $\rho(r)$ and $\Phi(r)$ satisfy all conditions introduced in
Section \ref{sec:ass}, can be applicable to the single-component
systems for any $\kappa r_{s} > 0$. It is important that from
Appendix \ref{sec:appsolutionn} it follows that {\it within the
basic model} (a1)-(a3) {\it the used alternative procedure is unique
one which provides that the solution $n(r)$ satisfies both the
neutrality condition} ({\ref{eq4}) {\it and the non-negativity
condition} (\ref{eq10}) {\it and and posses features} (b1) {\it and}
(b2).

Accordingly to Eqs.~(\ref{eq23}) and (\ref{eq26})-(\ref{eq27}), we
have always that $n(r=0) = 0$. In the case of the classical system
such a condition has to be satisfied from energetically reasons for
any $Z$ and $T$. However, in the case of degenerated system, when
the free particles have to be treated as quantum mechanical objects,
$n(r=0)$ is the small positive quantity. In this case the mentioned
condition practically  solve the problem of the elimination of DH
negative singularity, since provides the transition from
$n_{D}(r=0)=-\infty$ to $n(r=0)=0$.

From the expressions (\ref{eq29}), (\ref{eq23App}) and (\ref{eq30a})
one can see that the values of the potential $\Phi(r > r_{s})$ and
the probe particle potential energy $U$ are closed here to the
corresponding DH values $\Phi_{D}(r > r_{s})$ and $U_{\kappa} $ in
the region $\kappa r_{s} \ll 1$. This fact reflects the specificity
of the single-component systems where the screening constant
$\kappa$ is the same in both DH and presented methods. However, the
behavior of $\Phi(r)$ and $U$ essentially differs from the behavior
of $\Phi_{D}(r)$ and $U_{\kappa}$ in the single-component case for
large values of $\kappa r_{s}$. So, accordingly to Eq.~(\ref{eq25})
and (\ref{eqrpot2}), $U \approx U_{s}$ when $\kappa r_s \gg 1$ and
\begin{equation}
\label{eq35}  r_0 \cong r_s \cdot \left( {1 - \frac{1}{\kappa r_s }}
\right).
\end{equation}
Just such a case should realize in dusty plasmas containing dusty
particles with the charges $Ze$, where $|Ze| \gg 1$ and the
temperatures $T \approx 1000K$. The figure \ref{fig:rkrpot} suggests
that for $U$ it is suitable to use Eq.~(\ref{eq30a}) for $x < 3 /
2$, and Eq.~(\ref{eqUs}) for $x > 3/2$. The point $x=3/2$ can be
interpreted as an arbitrary border between the region of weak
non-ideality (DH region) and the region of strong non-ideality
(non-DH region).

One of the main results of the method which is developed in this
paper is appearing of three new parameters $r_{0}$, $\gamma_{s}$ and
$\gamma_{\kappa}$ which are given by Eqs.~(\ref{eq26}), (\ref{eq25})
and (\ref{eq27}). The parameter $r_{0}$ represents the radius of the
sphere centered in the probe particle which is classically forbidden
for the particles from their environment, while $\gamma_{s}$ and
$\gamma_{\kappa}$ can be interpreted as some kind of non-ideality
parameters. Full sense of these quantities will be discussed in Part
3. Also, in Part 3 the results of this paper will be used for
obtaining of other relevant screening parameters. Finally, let us
draw attention that the alternative procedure, described in this
paper, will be applied in Part 2 of this work in connection with
two-component systems.

\begin{acknowledgments}
The authors wish to thank to Prof. V.M. Adamyan for useful
discussion. The authors are thankful to the University P. et M.
Curie of Paris (France) for financial support, as well as to the
Ministry of Science of the Republic of Serbia for support within the
Project 141033 "Non-ideal laboratorial and ionospheric plasmas:
properties and applications".
\end{acknowledgments}

\begin{appendix}

\section{DH solutions}
\label{sec:appDebye}

\subsection{The procedure and expressions.}
\label{sec:pex}

In DH procedure the free particle density is expressed by means of
equation (\ref{eq9}) over the electrostatic potential and in such a
form used in Poisson's equation (\ref{eq5}). Then, by means of
(\ref{eq1}) and (\ref{eq2}), Eq. (\ref{eq5}) transforms to
Helmholtz's equation:
\begin{equation}\label{eqHelm}
\nabla^{2} \Phi (r) = \kappa ^2\Phi (r),
\end{equation}
which applies in the whole space $0 < r < \infty$, neglecting the
conditions (\ref{eqrrs}) and (\ref{eq12}). The solution $\Phi _D
(r)$ is determined by the boundary conditions (\ref{eq6}) and
(\ref{eq7}). After that, DH charge and particle densities
$\rho_{D}(r)$ and $n_{D}(r)$ are obtained in the whole region $r >
0$, by means of (\ref{eq2}) and (\ref{eq9}). These solutions are
given by
\begin{equation}\label{eqA1}
 \begin{array}{l} \displaystyle{\Phi_D (r) = Ze\cdot r^{-1}\exp
(- \kappa r)}, \\ \displaystyle{\rho_D (r) = -(Ze\kappa^2/4\pi)
\cdot r^{-1}\exp ( - \kappa r)}, \\ \displaystyle{n_D (r) = N -
(\kappa^2/4\pi)\cdot r^{-1}\exp (- \kappa r)},
\end{array}
\end{equation}
where $\kappa $ is given by (\ref{eq18}) or (\ref{eq19}). DH values
$\varphi _{D} $ and $U_{D}$ of the potential in the point $O$ and
the probe particle potential energy are determined from (\ref{eq7}),
(\ref{eq8}) and (\ref{eqA1}) and given by
\begin{equation}\label{eqA2}
 \varphi _{D} = - Ze \cdot \kappa, \qquad U_{D} = - (Ze)^{2}\cdot
\kappa \equiv -(Ze)^2/r_{\kappa},
\end{equation}
which are very often used in plasma physics. One can see that DH
value $U_{D}=U_{\kappa}$, where $U_{\kappa}$ denotes the boundary
value of the potential energy in the expression (\ref{eq30a}).

\subsection{The solution $n_D (r)$: the region $r < r_{s}$.} From
the expression (\ref{eqA2}) it follows that $n_D (r)<0$ in the
region $r < r_{in}^{( - )} $, where $r_{in}^{( - )} $ is the root of
the equation
\begin{equation}
n_{D}(r) = 0 ,
\label{eqnula1}
\end{equation}
which exists for any $\kappa > 0$. Then, the same expression for
$n_D (r)$ shows that the existing of the region where $n_D (r) < 0$
is compensated within a wider region $r < r_{out}^{( - )}$, where
$r_{out}^{( - )}$ is the root of the equation
\begin{equation} \int\limits_0^{r} {n_D
(r') \cdot 4\pi r'^2dr' = 0} ,
\label{eqnula2}
\end{equation}
which also exists for any $\kappa > 0$. From the neutrality
condition (\ref{eq4}) it follows that always $r_{out}^{( - )} <
r_{s}$. All mentioned is illustrated by Fig.~\ref{fig:nrhoD}.

\subsection{The solution $n_D (r)$: the region $r \ge r_{s}$.}
\label{sec:pex1}
We will take into account that in accordance with (\ref{eqA1}) and
(\ref{eq27}) the condition (\ref{eq12}) with $n(r) = n_D (r)$ can be
presented in the form
\begin{equation}\label{eqA3}
 h_D (r) = \frac{x^2}{3}\frac{\exp ( -xr/r_{s})}{r/r_{s}}
\ll 1, \qquad h_D (r) \equiv \frac{n_{D}(r)-N}{N}.
\end{equation}
From here it follows that the quantity $h_D (r)$ monotonously
increases when $\tilde {r}$ decreases, for each fixed $x.$
Therefore, in the region $r \ge r_s $, the quantity $h_D (r)$
reaches its maxima at $r = r_{s}$, which it means that it is enough
to consider the behavior of the quantity $h_D (r_s )$, which is
given by
\begin{equation}\label{eqA4}
 h_D (r_s ) = (x^2/3) \cdot \exp ( - x).
\end{equation}
as a function of $x$. From this expression one can see that $h_D
(r_s ) = 0$ at $x = 0$ and $x = \infty $, as well as that this
quantity reaches its maxima at $x = 2$. Since
\begin{equation}\label{eqA5}
 h_D (r_s )_{x = 2} = (4/3) \cdot e^{ - 2},
\end{equation}
we can consider that $n_D (r)$ satisfies the condition (\ref{eq12})
in the region $r > r_s $, for any $\kappa r_s > 0$. However, the
decreasing $h_{D}(r_{s})$, when $x$ increases in the region $x > 2$,
means that in this region the behavior of $n_{D}(r)$ becomes
non-physical in the whole space.

\section{The behavior of the solution $n(r)$}
\label{sec:appsolutionn}

\subsection{The region $r < r_{s}$: the extrapolation procedure.}
Because of the discussion in \ref{sec:nonD} in the region $r <
r_{s}$ is not possible using of any procedure based on the equation
(\ref{eqmu}), we have to return for a moment to the solution
$n_{D}(r)$ and consider it from the aspect of the possibility to
obtain it avoiding the procedure described in \ref{sec:pex}. Namely,
$n_{D}(r)$ can be treated as a result of an extrapolation of the
expression (\ref{eq22a}) in as wide as possible area ($0 < r <
\infty$) allowed by the condition (\ref{eq4}). Such a way of
extrapolation provides that the coefficient $a$ in (\ref{eq22a})
takes just DH value $\kappa^{2}/4\pi$. This procedure is consistent
one since $n_{D}(r)$, accordingly to Eq.~(\ref{eqA5}), satisfies the
condition (\ref{eq12}) in the region $r \ge r_s $ for any $\kappa
r_s > 0$.

The solutions obtained in a similar manner are already known in
physics. For example it is enough to mention Slater's and
Bates-Damgaard's wave functions which have successfully been used in
atomic physics \cite{sla30,bat49,gom50,sob79}. The applicability of
that solutions was caused by adequately chosen boundary conditions.
However, in the case of solution $n_D (r)$ such conditions were not
used, what caused its negativity in the region $0 < r <
r_{in}^{(-)}$.

In connection with this one should keep in mind that $n_{D}(r)$,
independently from its non-physicality in the region $0 < r <
r_{out}^{(-)}$, where $r_{in}^{(-)} < r_{out}^{(-)} < r_{s}$,
provides acceptability of the potential $\Phi_{D}(r)$, as well as
the screening characteristic length $r_{\kappa}$, for weakly and
moderately non-ideal single-component systems ($\kappa r_{s}
\lesssim 1$). This fact suggests that an adequate procedure of
extrapolation of the expression (\ref{eq22a}), which excludes
appearing of negative values of the solution $n(r)$, as well as its
non-physical behavior in the point $O$, would provide the
applicability of $n(r)$ not only to weakly and moderately non-ideal
systems, but to the systems of higher non-ideality ($\kappa r_{s}
\gg 1$).

Accordingly to above mentioned the procedure of obtaining of $n(r)$
has to be continued by the extrapolation of the expression
(\ref{eq22a}) in as wide as possible region of $r$, namely $r_{0} <
r < \infty$, allowed by the condition (\ref{eq10}). From
(\ref{eq22a}) it follows then that $r_{0}$ is the root of equation
\begin{equation}\label{eq22f}
N - a\cdot \exp ( - \kappa r)/r = 0,
\end{equation}
and consequently
\begin{equation}\label{eq22c}
 a = N \cdot r_{0} \exp (\kappa r_{0}).
\end{equation}
In the region $0 < r < r_{0}$ the solution $n(r)$ has to be
continued by means of equality
\begin{equation}\label{eq22d}
 n(r) \equiv 0, \qquad 0 < r < r_{0},
\end{equation}
which provides that this solution at least does not increases when
$r \to 0$. The corresponding form of the complete solution $n(r)$ is
given by the expression (\ref{eq23}). The parameter $r_{0}$ in this
expression has to be obtained from the neutrality condition in the
form (\ref{eq4a}). The same condition shows that it is always $r_{0}
< r_{s}$.

The complete procedure of obtaining of $n(r)$ is consistent one
since $n(r)$, given by (\ref{eq23}) and (\ref{eq26})-(\ref{eq27}),
satisfies the condition (\ref{eq12}) in the region $r \ge r_{s}$ for
any $\kappa r_{s} > 0$, as it follows from (\ref{eqB7}). Finally, it
is very important to draw one's attention that described
extrapolation procedure is unique one which provides that the
solution $n(r)$ is self-consistent and simultaneously satisfies all
conditions from Section \ref{sec:ass}, including the additional
condition (\ref{eq10}). Namely, as it can be shown, any other
extrapolation procedure causes the appearance of at least one
parameter that cannot be determined within that procedure itself.

\subsection{The region $r \ge r_{s}$.} From (\ref{eq1}),
(\ref{eq23}), (\ref{eq26}) and (\ref{eqA1}) it follows that the
product $N r_0 \exp (\kappa r_0 )$ in the expression for $n(r)$
differs from the coefficient $\kappa^2/4\pi$ in the expression for
$n_{D}(r)$ only by the factor $\chi (x)$. The behavior of $\chi
(x)$, given by (\ref{eq27}) and (\ref{eq29}), is presented in
Fig.~\ref{fig:chi}. One can see that in the region $0 < x \le 2$
this function increases from $1.0$ to a value that is close to
$1.2$. From this it follows that for $\kappa r_s \le 2$ the solution
$n(r)$ has to automatically satisfy the condition (\ref{eq12}). In
the region $\kappa r_s > 2$, where the non-physical properties of
$n_D (r)$ appear, we have to directly analyze the left side of
Eq.~(\ref{eq12}). From here and (\ref{eq23}) it follows that
\begin{equation}\label{eqB6}
 h (r_s ) = (x^2/3) \cdot \exp ( - x) \cdot \chi (x), \qquad
 h (r_s ) \equiv \frac{n(r_{s})-N}{N},
\end{equation}
and, for the difference of $h_{D}(r_{s})$, monotonously increases
whit the increasing of $x$ in the whole region $x > 0$. Then, on the
base of (\ref{eq29}) and (\ref{eqB6}), we obtained that
\begin{equation}\label{eqB7}
 \lim\limits_{x \to \infty } h(r_s ) = e^{ - 1}.
\end{equation}
Above mentioned means that $n(r)$ is free of the non-physical
properties, which have $n_{D}(r)$, and satisfies the condition
(\ref{eq12}) in the region $r \ge r_{s}$ for any $\kappa r_s > 0$.

\section{The solution $\Phi (r)$ and the potential $\varphi$}
\label{sec:appaccessory}

As it is known, the solution of Poisson's equation (\ref{eq5}),
which satisfies the boundary conditions (\ref{eq6}) and (\ref{eq7}),
is given by the expression
\begin{equation}\label{eqB1}
 \Phi (r) = \frac{1}{r}\left[ {Ze + \int\limits_0^r {\rho (r')}
4\pi r'^2dr'} \right] + \int\limits_r^\infty {\frac{\rho (r')}{r'}}
4\pi r'^2dr'.
\end{equation}
This expression, by means of the electro-neutrality condition
(\ref{eq4}), gets the form
\begin{equation}\nonumber
 \Phi (r) = - \frac{1}{r}\left[ {\int\limits_r^\infty {\rho (r')}
4\pi r'^2dr'} \right] + \int\limits_r^\infty {\frac{\rho (r')}{r'}}
4\pi r'^2dr',
\end{equation}
wherefrom it follows the expression
\begin{equation}\label{eqB3}
 \Phi (r) = - 4\pi \int\limits_r^\infty {\rho (r')\left(
{\frac{1}{r} - \frac{1}{r'}} \right)} r'^2dr'.
\end{equation}
That is one out of the two expressions for the potential $\Phi (r)$
which are used within the frame of this work. The basic feature of
(\ref{eqB3}) is that for each $r > 0$ the potential $\Phi (r)$ is
expressed only by $\rho(r')$ from the region $r' > r$.

From the definition (\ref{eq7}) of the potential $\varphi $ and
(\ref{eqB1}) it follows the expression
\begin{equation}\label{eqB2}
 \varphi = \int\limits_0^\infty {\frac{\rho (r)}{r}} 4\pi r^2dr,
\end{equation}
which uses in this work for the determination of  $\varphi $.

Finally, by means of (\ref{eqB1}) and (\ref{eqB2}) one obtains
another expression for the potential $\Phi (r)$, namely
\begin{equation}\label{eqB4}
 \Phi (r) = \frac{Ze}{r} + \varphi - 4\pi \int\limits_0^r {\rho
(r')\left( {\frac{1}{r'} - \frac{1}{r}} \right)} r'^2dr'.
\end{equation}
The basic feature of (\ref{eqB4}) is that for each different $r =
r_1 $ and$r = r_2 $, the difference $\left[ {\Phi (r_2 ) - \Phi (r_1
)} \right]$ is expressed only by $\rho (r')$ in the region $0 < r' <
\max \left( {r_1 ,r_2 } \right)$. This fact will play a significant
role in Part 2.

\end{appendix}

\newcommand{\noopsort}[1]{} \newcommand{\printfirst}[2]{#1}
  \newcommand{\singleletter}[1]{#1} \newcommand{\switchargs}[2]{#2#1}

\end{document}